\documentclass[
  a4paper,
  amsfonts,
  amssymb,
  amsmath,
  reprint,
  floatfix,
  nofootinbib,
  pra,
  aps,
  superscriptaddress
]{revtex4-2}

\usepackage[english]{babel}
\usepackage[T1]{fontenc}

\usepackage{physics}

\usepackage{graphicx}
\usepackage{booktabs}
\usepackage{multirow}

\usepackage{hyperref}
\hypersetup{
    colorlinks=true,
    linkcolor=blue,
    citecolor=blue,
    urlcolor=blue
}

\begin{document}
\title{A Divide-and-Conquer Quantum-Selected Configuration
Interaction for Evaluating $\bf \pi$--$\bf \pi$ Stacking Interaction Energies in the Benzene Dimer}

\author{Ryotaro Tajima}
\thanks{These authors contributed equally to this work.}
\affiliation{Shionogi \& Co., Ltd.,
  5-54 Ofukacho, Kita-ku, Osaka, 530-0011, Japan}
\email[Correspondence email address: ]{ryotaro.tajima@shionogi.co.jp}

\author{Rei Sato}
\thanks{These authors contributed equally to this work.}
\affiliation{Classiq Technologies G.K.,
  1-5-1 Marunouchi, Chiyoda-ku, Tokyo, 100-6509, Japan}
\email[Correspondence email address: ]{rei@classiq.io}

\author{Yosuke Iyama}
\affiliation{Shionogi \& Co., Ltd.,
  5-54 Ofukacho, Kita-ku, Osaka, 530-0011, Japan}
\email[Correspondence email address: ]{yosuke.iyama@shionogi.co.jp}

\author{Ryo Kiguchi}
\affiliation{Shionogi \& Co., Ltd.,
  5-54 Ofukacho, Kita-ku, Osaka, 530-0011, Japan}
\email[Correspondence email address: ]{ryo.kiguchi@shionogi.co.jp}

\author{Yoshitake Kitanishi}
\affiliation{Shionogi \& Co., Ltd.,
  5-54 Ofukacho, Kita-ku, Osaka, 530-0011, Japan}
\email[Correspondence email address: ]{yoshitake.kitanishi@shionogi.co.jp}


\begin{abstract}
    Accurate evaluation of $\pi$--$\pi$ stacking interactions requires a description of electron correlation on a small energy scale. Direct variational quantum calculations of the benzene dimer with CAS(28e,20o) require 40 qubits and deep circuits. Here, we propose Divide-and-Conquer Quantum-Selected Configuration Interaction (Deep QSCI), combining the divide-and-conquer idea of Deep VQE with QSCI, and apply it to the sandwich benzene dimer. We perform QSCI for one monomer and construct a reduced local basis by applying particle-number-conserving one-electron excitations to its ground state. Using this basis and the intermonomer interaction Hamiltonian, we build and diagonalize an effective Hamiltonian of the dimer. Since the monomers have the same structure, the QSCI result is reused for both monomers and across intermolecular distances, reducing the required quantum register from 40 to 20 qubits for the monomer CAS(14e,10o). With 6-31G**, the model gives an attractive minimum of -0.91 kcal/mol at 4.0 \AA, compared with the counterpoise-corrected CCSD(T) value of -1.14 kcal/mol. With cc-pVDZ, a large cancellation appears between the product-state interaction energy and the energy lowering by diagonalization. This may reflect insufficient consistency of the active orbitals and reduced local bases between monomers. Orbital consistency, reduced-space convergence across basis sets, and charge transfer remain key issues for quantitative accuracy. Deep QSCI thus provides a resource-efficient approach to constructing model interaction curves without repeating monomer quantum calculations at each separation.
\end{abstract}

\maketitle

\section{Introduction}
\label{sec:introduction}
Despite remarkable advances in modern medicine, substantial unmet medical needs remain across many disease areas~\cite{sharp2026quantify}.  Therefore, efficient and reliable methods for drug discovery are still needed. The development of new drugs remains a costly and high-risk endeavor, with low success rates throughout the development pipeline~\cite{waring2015analysis, scannell2012diagnosing}. Numerous candidate compounds fail during preclinical and clinical development, resulting in substantial attrition and increasing research and development costs~\cite{hay2014clinical}.

Therefore, improving the accuracy and efficiency of early-stage drug discovery remains a major challenge. In particular, the identification and optimization of candidate compounds require reliable prediction of intermolecular interactions~\cite{schneider2018automating,niazi2025quantum}, which play a central role in determining molecular recognition and binding affinity. Accurate evaluation of these interactions is therefore essential for rational drug design.

Among such intermolecular interactions, $\pi$-$\pi$ stacking between aromatic rings plays an important role in stabilizing protein–ligand complexes and controlling their binding affinity, selectivity, and conformation.  An accurate description of $\pi$-$\pi$ stacking is therefore important for the reliable evaluation of molecular binding and for rational drug design. Dispersion forces, electrostatic interactions, and electron correlation all contribute to this interaction. Their complex interplay makes its quantitative description a major challenge in electronic-structure calculations.

Quantum computing has been studied as a promising approach for calculating molecular electronic states, including electron correlation, and may provide a way to describe weak intermolecular interactions with high accuracy. Variational quantum algorithms, such as the variational quantum eigensolver (VQE), have been widely applied to ground-state energy calculations for molecular Hamiltonians, including those of H$_2$ and LiH~\cite{peruzzo2014variational, kandala2017hardware}.  More recently, quantum-selected configuration interaction (QSCI) has been proposed as a quantum–classical hybrid method~\cite{qsci-kannno}.  In QSCI, important electronic configurations are sampled from a wave function prepared on a quantum computer. The Hamiltonian projected onto the selected subspace is then diagonalized on a classical computer. Sampling-based approaches built on QSCI combine quantum computation with high-performance classical computing and have extended the range of accessible active-space electronic-structure problems~\cite{robledo2025chemistry}.

Quantum computing has also been applied to the evaluation of intermolecular interactions. For noncovalent interactions such as $\pi$-$\pi$ stacking, the interaction energy is obtained as a small difference between the energy of the combined system and those of its isolated components. Errors in the individual electronic energies therefore directly affect the accuracy of the interaction energy. Anderson et al. evaluated London dispersion interactions on a quantum processor using a quantum Drude oscillator model and a variational quantum algorithm~\cite{PhysRevA.105.062409}.  Kaliakin et al. combined a sampling-based quantum–classical approach with the supermolecular method to calculate potential-energy surfaces and interaction energies for the water and methane dimers on quantum processors~\cite{kaliakin2025accurate}.  Near the equilibrium geometries, their results agreed well with active-space CASCI calculations.  Deviations from full-basis CCSD(T) results were also reported to be within $1$ kcal/mol.  

These studies show that applications of quantum computing are expanding from electronic-structure calculations for isolated molecules to the evaluation of weak noncovalent interactions. However, only a limited number of studies have applied quantum computing to $\pi$-$\pi$ stacking between aromatic molecules. The benzene dimer is a representative model system for this interaction.  The stable structures and interaction energies of the benzene dimer have been extensively studied using various classical electronic-structure methods, including high-level coupled-cluster and quantum Monte Carlo approaches~\cite{miliordos2014benchmark,azadi2015chemical}.

From the viewpoint of variational quantum computation, VQE calculations of the ground-state energy of a benzene monomer have been reported~\cite{sennane2023calculating}.  A direct treatment of all orbitals of benzene in the STO-3G basis requires $72$ qubits. Previous studies have therefore used active spaces requiring only $8$–$16$ qubits~\cite{sennane2023calculating}.  These studies have also identified difficulties in variational optimization as the active space becomes larger, together with errors caused by quantum-device noise~\cite{sennane2023calculating}.  A direct all-orbital treatment of the benzene dimer in the same basis would nominally require $144$ qubits.  A scalable method is therefore needed to limit the growth of quantum resources while retaining the electronic degrees of freedom required to describe $\pi$-$\pi$ stacking accurately.  Such a method must also provide sufficiently accurate interaction energies.  In addition, variational quantum algorithms such as VQE can suffer from trainability problems, including barren plateaus, as the system size and quantum circuit depth increase~\cite{mcclean2018barren}.

In this work, we propose Divide-and-Conquer Quantum-Selected Configuration Interaction (Deep QSCI) that combines the divide-and-conquer framework of Deep VQE with the subspace-selection procedure of QSCI.  Deep VQE, which forms the basis of our method, introduces a divide-and-conquer strategy into VQE to reduce the scaling of quantum resources~\cite{Fujii2022DeepVQE}.  It considers a large Hamiltonian composed of subsystem Hamiltonians and interaction terms that couple the subsystems.  A reduced space is constructed from the low-energy states of each subsystem, and an effective Hamiltonian for the full system is then built in this space.  This procedure reduces the required number of qubits and quantum-circuit resources.  In contrast, QSCI constructs a subspace from important electronic configurations obtained by measuring a quantum state.  The Hamiltonian projected onto this subspace is diagonalized classically.  QSCI can therefore reduce the dependence on large-scale optimization of variational parameters~\cite{qsci-kannno}.

Deep QSCI combines these two features.  For the identical rigid monomers considered here, one isolated-monomer QSCI workflow is performed for each choice of basis set and active space.  The resulting state and reduced local basis are reused for both monomers throughout the intermolecular-separation scan.  Only the geometry-dependent interaction and the dimer effective Hamiltonian are evaluated at each separation, and these steps are performed classically.  We calculate the ground-state energy of the benzene-dimer model from this effective Hamiltonian and evaluate the $\pi$-$\pi$ stacking interaction energy from the difference between the dimer and isolated-monomer energies.  Through these calculations, we examine whether this reuse-based construction can produce physically meaningful model interaction-energy curves while reducing the required quantum register.

The remainder of this paper is organized as follows.  In Sec.~\ref{sec:benzene dimer}, we present the computational settings for the benzene dimer and the procedure used to evaluate its interaction energy.  In Sec.~\ref{sec:preliminaries}, we describe the theoretical frameworks of QSCI and Deep VQE.  In Sec.~\ref{sec:deep_qsci}, we introduce the framework of Deep QSCI.  In Sec.~\ref{sec:computational_setup}, we explain the computational setup used to evaluate Deep QSCI.  In Sec.~\ref{sec:result}, we report the ground-state and $\pi$-$\pi$ stacking interaction energies and compare them with classical electronic-structure results.  In Sec.~\ref{sec:discussion}, we discuss the scope and limitations of the present model.  Finally, Sec.~\ref{sec:conclusion} summarizes our conclusions.

\section{Benzene dimer}
\label{sec:benzene dimer}
\begin{figure}
    \centering
    \includegraphics[width=1.0\linewidth]{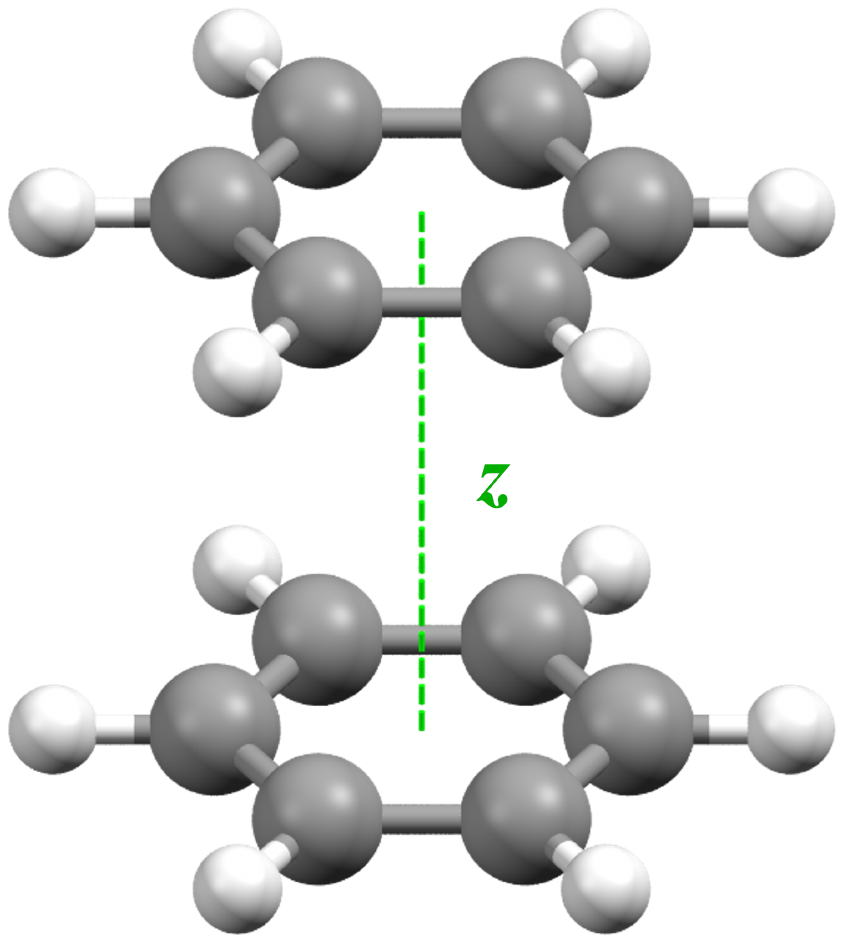}
    \caption{Sandwich-type benzene dimer molecule.  $z$ is the intermolecular separation between the two monomers.}
    \label{fig:benzene dimer}
\end{figure}

The benzene dimer is a noncovalently bound molecular complex consisting of two benzene molecules and is widely studied as a prototypical system for aromatic $\pi$–$\pi$ interactions.  Although the molecular structure of each benzene monomer is relatively simple, the intermolecular interaction is weak, with a binding energy of only a few $\mathrm{kcal/mol}$.  The benzene dimer has therefore been extensively investigated both theoretically and experimentally as a representative system for weak intermolecular interactions.

Three representative configurations are commonly considered: the sandwich, $T$-shaped, and parallel-displaced configurations.  In the sandwich configuration shown as Fig.~\ref{fig:benzene dimer}, the two benzene rings are arranged parallel to each other with their centers aligned along a common axis.  In the $T$-shaped configuration, the two rings are approximately perpendicular to each other, whereas in the parallel-displaced configuration they remain approximately parallel but are laterally displaced.  High-level electronic-structure calculations have been performed for these configurations to characterize their intermolecular potential energy surfaces.

For the sandwich configuration, Sinnokrot and Sherrill investigated the interaction energy using second-order M{\o}ller--Plesset perturbation theory (MP2) and coupled-cluster theory with single, double, and perturbative triple excitations [CCSD(T)] in combination with augmented correlation-consistent basis sets. Their estimated CCSD(T)/aug-cc-pVQZ* potential energy curve yields an equilibrium intermolecular separation of approximately $z=3.9~\text{\AA}$, with a corresponding interaction energy of approximately $E_{\mathrm{int}}=-1.70~\mathrm{kcal/mol}$~\cite{sinnokrot2004highly}.

An estimated complete-basis-set CCSD(T) interaction energy of $E_{\mathrm{int}}=-1.81~\mathrm{kcal/mol}$ has also been reported~\cite{sinnokrot2004highly}.  Consistently, Wheeler reported a CCSD(T) interaction energy of $-1.79~\mathrm{kcal/mol}$ for the benzene sandwich dimer~\cite{wheeler2011local}. Therefore, these high-level coupled-cluster results place the interaction energy of the sandwich benzene dimer near $-1.8~\mathrm{kcal/mol}$.

In this work, we use the sandwich configuration of the benzene dimer as a benchmark for QSCI and Deep QSCI.

\section{Preliminaries}
\label{sec:preliminaries}
In this section, we review the theoretical foundations of QSCI and Deep VQE required for the subsequent discussion.

\subsection{QSCI}
\label{subsec:qsci}

QSCI is a hybrid quantum--classical method in which a quantum processor is used to identify important electronic configurations and the Hamiltonian projected onto the
space spanned by those configurations is diagonalized on a classical computer~\cite{qsci-kannno}.  For a ground-state calculation, an input state $\ket{\psi_{\mathrm{in}}}$ that approximately represents the target state is first prepared on a quantum computer. The input state need not reproduce all CI coefficients accurately, but it should contain the configurations that are important for the target state.  In the calculations reported below, the direct dimer QSCI calculation uses a unitary coupled-cluster ansatz with single excitations (UCCS), whereas the monomer QSCI calculation used within Deep QSCI employs single and double excitations (UCCSD); further details are given in Sec.~\ref{sec:computational_setup}.

The input state is measured $N_{\mathrm{shot}}$ times in the computational basis.  The probability of observing a bit string $x$ is
\begin{equation}
  P(x) = \left|
      \left\langle x\,\middle|\,\psi_{\mathrm{in}}\right\rangle
    \right|^2.
  \label{eq:qsci_sampling_probability}
\end{equation}
The $R$ most frequently observed computational-basis states are retained to form the set
\begin{equation}
  \mathcal{S}_R
  = \left\{
  \ket{x}\;\middle|\;
  x\in\{0,1\}^{N_q},\;
  x \in R\right\},
  \label{eq:qsci_selected_set}
\end{equation}
where $N_q$ is the number of qubits and $R$ is a user-defined positive integer~\cite{qsci-kannno}.  The value of $R$ controls the trade-off between the quality of the selected space and the cost of the classical diagonalization.

The Hamiltonian is then diagonalized in the subspace spanned by $\mathcal{S}_R$. The corresponding eigenvalue problem is
\begin{equation}
  H_R\boldsymbol{c}=E_R\boldsymbol{c},
  \label{eq:qsci_eigenvalue_problem}
\end{equation}
where $H_R$ is the $R\times R$ Hermitian matrix defined by
\begin{equation}
  (H_R)_{xy}=\mel{x}{\hat{H}}{y},
  \qquad \ket{x},\ket{y}\in\mathcal{S}_R.
  \label{eq:qsci_projected_hamiltonian}
\end{equation}
The matrix elements and the diagonalization are evaluated classically. The smallest eigenvalue $E_R$ gives the QSCI approximation to the ground-state energy, and its normalized eigenvector provides the coefficients of the
output state
\begin{equation}
  \ket{\psi_{\mathrm{out}}}
  =\sum_{\ket{x}\in\mathcal{S}_R}c_x\ket{x}.
  \label{eq:qsci_output_state}
\end{equation}

Since $H_R$ is diagonalized exactly within the selected subspace, the Rayleigh--Ritz variational principle yields
\begin{equation}
  E_{\mathrm{exact}}\leq E_R,
  \label{eq:qsci_variational_bound}
\end{equation}
where $E_{\mathrm{exact}}$ is the exact ground-state energy of $\hat{H}$ in the symmetry sector under consideration.  


\subsection{Deep VQE}
\label{subsec:deep_vqe}

Deep VQE is a divide-and-conquer method that approximates the low-energy sector of a large system using quantum processors with fewer qubits~\cite{Fujii2022DeepVQE}.  Although the method can be extended recursively to multiple levels, we focus here on its basic two-stage formulation. 

The system is divided into $N$ subsystems, and the Hamiltonian is written as
\begin{equation}
  \hat{H}=\sum_i\hat{H}_i+\sum_{ij}\hat{V}_{ij},
  \label{eq:deep_vqe_total_hamiltonian}
\end{equation}
where $\hat{H}_i$ acts only on subsystem $i$ and $\hat{V}_{ij}$ couples subsystems $i$ and $j$. The interaction can be decomposed into products of local operators as
\begin{equation}
  \hat{V}_{ij}
  =\sum_{\nu}v_{\nu}
  \hat{W}_{\nu}^{(i)}\otimes\hat{W}_{\nu}^{(j)}.
  \label{eq:deep_vqe_interaction_decomposition}
\end{equation}

Deep VQE first applies VQE independently to each subsystem while neglecting the intersubsystem interactions. For a subsystem containing $n$ qubits, this step prepares an approximate local ground state
\begin{equation}
  \ket{\psi_0^{(i)}}
  =\hat{U}_i\!\left(\boldsymbol{\theta}^{(i),*}\right)\ket{0^n}.
  \label{eq:deep_vqe_local_ground_state},
\end{equation}
where $\boldsymbol{\theta}^{(i),*}$ denotes the optimized variational parameters obtained by minimizing $\bra{\psi_0^{(i)}}\hat{H}_i\ket{\psi_0^{(i)}}$.

We then generate a $K$-dimensional local basis by applying local excitation operators to this state,
\begin{equation}
  \ket{\psi_k^{(i)}}
  =\hat{W}_k^{(i)}\ket{\psi_0^{(i)}},
  \label{eq:deep_vqe_local_basis}
\end{equation}
where $k=\{1,\ldots,K\}$ and $\hat{W}_1^{(i)}=\hat{I}$.  In the original construction, the remaining operators are chosen to represent excitations near the subsystem boundary that enter the intersubsystem interaction.

The states in Eq.~\eqref{eq:deep_vqe_local_basis} are generally nonorthogonal.  Their overlaps are measured on the subsystem quantum processor, and an orthonormal local basis is constructed through a linear transformation,
\begin{equation}
  \ket{\widetilde{\psi}_k^{(i)}}
  =\sum_{k'=1}^{K}P_{kk'}^{(i)}\ket{\psi_{k'}^{(i)}},
  \label{eq:deep_vqe_orthonormal_basis}
\end{equation}
where $P^{(i)}$ is obtained from the overlap matrix
$S_{kl}^{(i)}=\left\langle\psi_k^{(i)}\middle|\psi_l^{(i)}\right\rangle$.  Linearly dependent directions must be removed before this transformation is applied.

The subsystem and intersubsystem contributions to the effective Hamiltonian
are defined in the orthonormal local bases by
\begin{align}
  \left(H_i^{\mathrm{eff}}\right)_{kl}
  &=\mel{\widetilde{\psi}_k^{(i)}}{\hat{H}_i}
  {\widetilde{\psi}_l^{(i)}},
  \label{eq:deep_vqe_effective_local_hamiltonian}\\
  \left(V_{ij}^{\mathrm{eff}}\right)_{kk'll'}
  &=\sum_{\nu}v_{\nu}
  \mel{\widetilde{\psi}_k^{(i)}}{\hat{W}_{\nu}^{(i)}}
  {\widetilde{\psi}_l^{(i)}}
  \mel{\widetilde{\psi}_{k'}^{(j)}}{\hat{W}_{\nu}^{(j)}}
  {\widetilde{\psi}_{l'}^{(j)}}.
  \label{eq:deep_vqe_effective_interaction}
\end{align}
The full effective Hamiltonian is therefore
\begin{equation}
  \hat{H}^{\mathrm{eff}}
  =\sum_i\hat{H}_i^{\mathrm{eff}}
  +\sum_{ij}\hat{V}_{ij}^{\mathrm{eff}}.
  \label{eq:deep_vqe_effective_hamiltonian}
\end{equation}
For $N$ subsystems with the same local dimension $K$, this Hamiltonian acts on a $K^N$-dimensional reduced space and can be encoded using $m=N\left\lceil\log_2K\right\rceil$ qubits, rather than the $M=nN$ qubits required for the unreduced system.  A qubit reduction is obtained when $\lceil\log_2K\rceil<n$. 

We then perform a second VQE to minimize the expectation value of the effective Hamiltonian,
\begin{equation}
E^{\mathrm{eff}}(\boldsymbol{\phi})
= \mel{0^m}{
\hat{V}^{\dagger}(\boldsymbol{\phi})
\hat{H}^{\mathrm{eff}}
\hat{V}(\boldsymbol{\phi})
}{0^m},
\label{eq:deep_vqe_second_vqe}
\end{equation}
where $\hat{V}(\boldsymbol{\phi})$ is a parameterized quantum circuit acting on the reduced $m$-qubit space.  The results provide an approximation to the ground-state energy of the original Hamiltonian, while the optimized circuit represents the corresponding state within the reduced product space.

The accuracy of Deep VQE depends on the choice of subsystem partition and local basis~\cite{Mizuta2021DeepVQEexcited,Erhart2022LocalBases}.  The method is expected to perform well when the dominant correlations are captured within each subsystem and the selected local excitations adequately describe the intersubsystem coupling.  In the present application, the two benzene monomers are treated as separate subsystems.  Because the benzene dimer consists of two monomers coupled through a relatively weak noncovalent interaction, this decomposition provides a natural setting for applying Deep VQE.  In the following sections, we assess this construction by evaluating the resulting interaction-energy curves for the selected local spaces.

\section{Proposed method}
\label{sec:deep_qsci}
\begin{figure*}
    \centering
    \includegraphics[width=1.0\linewidth]{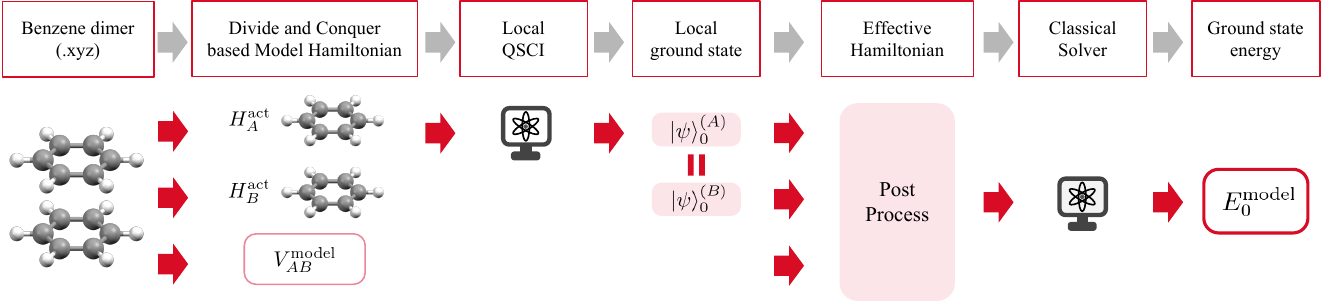}
    \caption{Schematic overview of the proposed Deep QSCI framework for the benzene dimer.  The dimer model Hamiltonian comprises the active-space monomer Hamiltonians, $\hat H_A^{\mathrm{act}}$ and $\hat H_B^{\mathrm{act}}$, and the model intermonomer interaction $\hat V_{AB}^{\mathrm{model}}$.  QSCI is performed once for an isolated monomer for each basis-set and active-space choice.  The resulting state is reused to construct the reduced local spaces for both $A$ and $B$.  These local spaces and $\hat V_{AB}^{\mathrm{model}}$ define an effective Hamiltonian in the reduced product space, which is diagonalized classically to obtain $E_{AB}^{\mathrm{Deep\,QSCI}}$.
    }
    \label{fig:placeholder}
\end{figure*}

Deep QSCI combines the subsystem construction of Deep VQE with the
configuration-selection procedure of QSCI. In the present application, the benzene dimer is divided into two identical monomers, denoted by $A$ and $B$.  For each choice of basis set and active space, one isolated-monomer QSCI workflow, including the scan over the state-preparation parameter, is performed.  The resulting multiconfigurational state and reduced local basis are reused for both monomers and at all separation.  To retain this reuse, we describe the two subsystems using the same isolated-monomer active-space Hamiltonian and introduce their geometry-dependent coupling separately. The resulting Hamiltonian is intended as a fragment-based model Hamiltonian and not as an exact decomposition of the full dimer active-space Hamiltonian.  The model Hamiltonian is represented and diagonalized classically in the product of the two reduced local spaces.

\subsection{Model Hamiltonian}
\label{subsec:deep_qsci_model_hamiltonian}
\begin{figure}
    \centering
    \includegraphics[width=1.0\linewidth]{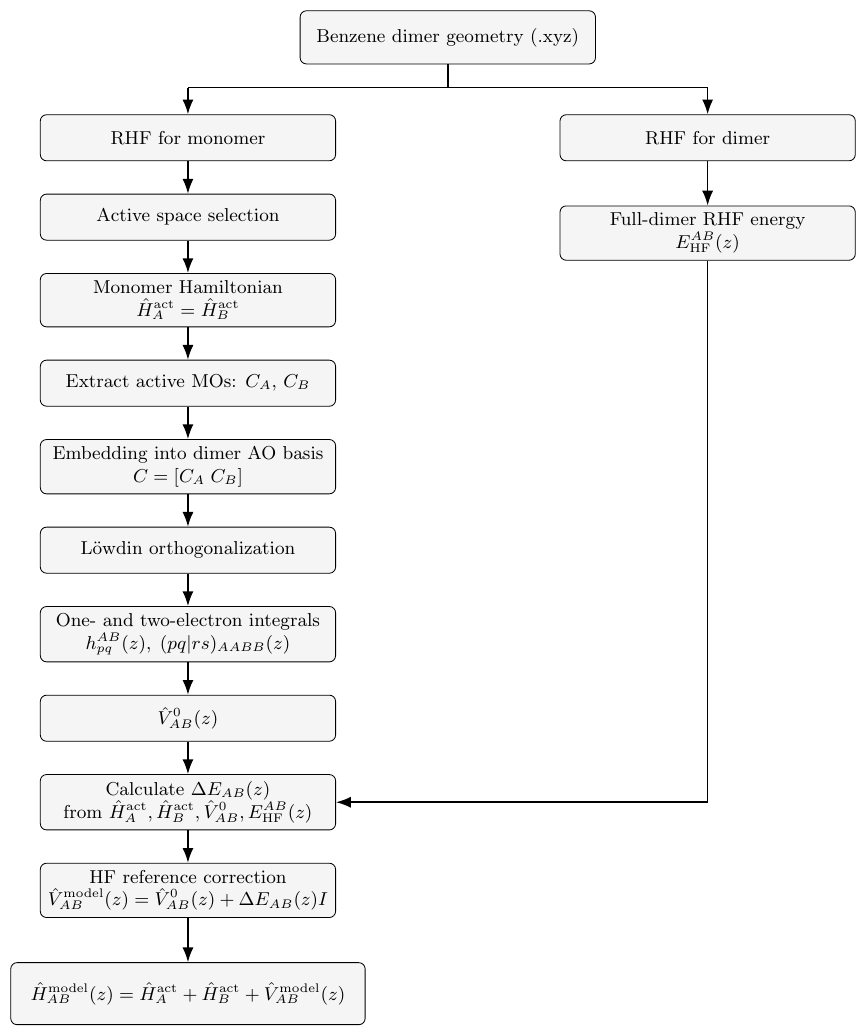}
    \caption{Schematic workflow for constructing the model Hamiltonian of the benzene dimer.  The monomer active-space Hamiltonians are obtained from isolated-monomer RHF calculations, while the model intermonomer interaction is constructed from L\"owdin-orthogonalized active orbitals with an HF reference correction.
    }
    \label{fig:flow-chart-model-hamiltonian}
\end{figure}

Let $\mathcal{A}_A$ and $\mathcal{A}_B$ denote the active orbital sets of the isolated monomers.  We introduce a model orbital space labeled by $\mathcal{A}_{AB}^{\mathrm{model}}=\mathcal{A}_A\oplus\mathcal{A}_B.$  For the extended active space used in this work, each monomer is described by $\mathrm{CAS}(14e,10o)$.  The corresponding model contains $28$ electrons in $20$ spatial orbitals.  In the calculations reported below, however, each monomer is restricted to its neutral sector with 14 electrons.  The resulting product space is therefore a subspace of $\mathrm{CAS}(28e,20o)$ and does not contain intermonomer charge-transfer configurations.

The local terms are copies of the isolated-monomer active-space Hamiltonian,
\begin{equation}
  \hat H_A^{\mathrm{act}}
  =\hat H_B^{\mathrm{act}}
  =\hat H_{\mathrm{mono}}^{\mathrm{act}}.
  \label{eq:deep_qsci_local_hamiltonians}
\end{equation}
The dimer model Hamiltonian at intermolecular separation $z$ is defined as
\begin{equation}
  \hat H_{AB}^{\mathrm{model}}(z)
  =\hat H_A^{\mathrm{act}}
  +\hat H_B^{\mathrm{act}}
  +\hat V_{AB}^{\mathrm{model}}(z).
  \label{eq:deep_qsci_model_hamiltonian}
\end{equation}
The interaction in Eq.~\eqref{eq:deep_qsci_model_hamiltonian} is constructed from selected one- and two-electron integrals evaluated at the dimer geometry.  It is introduced to couple the two identical isolated-monomer problems while preserving the reuse of their common QSCI local space. It is not obtained by subtracting the two isolated-monomer Hamiltonians from a full dimer
active-space Hamiltonian.

To evaluate these interaction integrals, the isolated-monomer active orbitals are first embedded into the atomic-orbital (AO) basis of the dimer. Let
\begin{equation}
  C=\begin{pmatrix}C_A&C_B\end{pmatrix}
  \label{eq:deep_qsci_embedded_orbitals}
\end{equation}
be the combined coefficient matrix after this embedding, and let
$S_{\mathrm{AO}}^{AB}$ be the dimer AO overlap matrix. The overlap matrix in the combined active orbital set is
\begin{equation}
  S_{\mathrm{MO}}
  =C^{\dagger}S_{\mathrm{AO}}^{AB}C.
  \label{eq:deep_qsci_mo_overlap}
\end{equation}
We apply symmetric L\"owdin orthogonalization,
\begin{eqnarray}
  \widetilde C&=&C S_{\mathrm{MO}}^{-1/2}\\
  \widetilde C^{\dagger}S_{\mathrm{AO}}^{AB}\widetilde C&=&I,
  \label{eq:deep_qsci_lowdin_orthogonalization}
\end{eqnarray}
and retain the original column ordering to write
\begin{equation}
  \widetilde C
  =\begin{pmatrix}\widetilde C_A&\widetilde C_B\end{pmatrix}.
  \label{eq:deep_qsci_orthogonalized_partition}
\end{equation}
The labels $A$ and $B$ in Eq.~\eqref{eq:deep_qsci_orthogonalized_partition}
are inherited from the corresponding columns before orthogonalization.  The combined orbital set is orthonormal, including between the two labeled groups.  Because the L\"owdin orthogonalization mixes the embedded orbitals, these labels do not imply that every orthogonalized orbital has support on only one
monomer.

Let $\{\widetilde\phi_p^A\}$ and $\{\widetilde\phi_q^B\}$ denote the orbitals defined by $\widetilde C_A$ and $\widetilde C_B$, respectively. The intermonomer one-electron integrals are
\begin{equation}
  h_{pq}^{AB}(z)
  =\left\langle
  \widetilde\phi_p^A
  \middle|
  \hat h_{\mathrm{core}}^{AB}(z)
  \middle|
  \widetilde\phi_q^B
  \right\rangle,
  \label{eq:deep_qsci_intermonomer_one_electron_integrals}
\end{equation}
where $\hat h_{\mathrm{core}}^{AB}(z)$ contains the electronic kinetic energy and the attraction to all nuclei in the dimer. The retained two-electron integrals are
\begin{align}
  (pq|rs)_{AABB}(z)
  =&\iint
  \widetilde\phi_p^{A*}(\boldsymbol r_1)
  \widetilde\phi_q^A(\boldsymbol r_1)
  \frac{1}{r_{12}}\notag\\
  &\times
  \widetilde\phi_r^{B*}(\boldsymbol r_2)
  \widetilde\phi_s^B(\boldsymbol r_2)
  \,d\boldsymbol r_1\,d\boldsymbol r_2.
  \label{eq:deep_qsci_intermonomer_two_electron_integrals}
\end{align}
Using these coefficients, we define
\begin{align}
  \hat V_{AB}^{0}(z)
  &=\sum_{pq,\sigma}
  (
  h_{pq}^{AB}(z)
  \hat a_{p\sigma,A}^{\dagger}\hat a_{q\sigma,B}
  +h_{pq}^{AB}(z)^*
  \hat a_{q\sigma,B}^{\dagger}\hat a_{p\sigma,A}
  )\notag\\
  &+\sum_{pqrs}\sum_{\sigma\tau}
  (pq|rs)_{AABB}(z)
  \hat a_{p\sigma,A}^{\dagger}
  \hat a_{r\tau,B}^{\dagger}
  \hat a_{s\tau,B}
  \hat a_{q\sigma,A}.
  \label{eq:deep_qsci_bare_model_interaction}
\end{align}
Here, $p,q$ label spatial orbitals assigned to $A$, $r,s$ label spatial orbitals assigned to $B$, and $\sigma,\tau\in\{\alpha,\beta\}$.  Since the pairs of electrons of the intermonomer are summed only once, with $p,q\in A$ and $r,s\in B$, no factor of $1/2$ is required for the two-electron term.  Equation~\eqref{eq:deep_qsci_bare_model_interaction} retains the intermonomer one-electron hopping terms and the $(AA|BB)$ Coulomb terms.  Other mixed two-electron integral classes are omitted.

The creation and annihilation operators in Eq.~\eqref{eq:deep_qsci_bare_model_interaction} act on the model Fock space defined by the isolated-monomer orbital indices. The interaction coefficients, however, are evaluated using the L\"owdin-orthogonalized orbitals and are assigned to the corresponding monomer indices by their column ordering. The isolated-monomer Hamiltonians and QSCI states are not transformed to the L\"owdin orthogonalized basis states.  Consequently, Eq.~\eqref{eq:deep_qsci_model_hamiltonian} is an index-mapped model Hamiltonian rather than an exact projection of the full dimer Hamiltonian onto a common orthonormal orbital basis. This index mapping is an additional approximation of the present method.

To align the energy reference of the reduced interaction model with that of the full dimer, we introduce a geometry-dependent HF reference correction. Specifically, $\Delta E_{AB}(z)$ is defined as the difference between the full-dimer RHF energy and the expectation value of the uncorrected model Hamiltonian, $\hat H_A^{\mathrm{act}}+\hat H_B^{\mathrm{act}}+\hat V_{AB}^{0}(z)$, evaluated for the product of the monomer HF determinants.  The corrected interaction operator is then defined as
\begin{equation}
  \hat V_{AB}^{\mathrm{model}}(z)
  =\hat V_{AB}^{0}(z)+\Delta E_{AB}(z)\hat I.
  \label{eq:deep_qsci_corrected_model_interaction}
\end{equation}
Let
\begin{equation}
  \ket{\Phi_{\mathrm{HF}}^{AB}}
  =\ket{\Phi_{\mathrm{HF}}^A}
  \otimes\ket{\Phi_{\mathrm{HF}}^B}
  \label{eq:deep_qsci_hf_product_state}
\end{equation}
be the product of the isolated-monomer HF states in the model Fock space. The correction is chosen as
\begin{align}
  \Delta E_{AB}(z)
  &={}E_{\mathrm{HF}}^{AB}(z)\notag\\
  &-\bra{\Phi_{\mathrm{HF}}^{AB}}
  (\hat H_A^{\mathrm{act}}
  +\hat H_B^{\mathrm{act}}
  +\hat V_{AB}^{0}(z)
  )
  \ket{\Phi_{\mathrm{HF}}^{AB}}
  \label{eq:deep_qsci_hf_reference_correction}
\end{align}
where $E_{\mathrm{HF}}^{AB}(z)$ is the full-dimer HF energy. It follows that
\begin{equation}
  \bra{\Phi_{\mathrm{HF}}^{AB}}
  \hat H_{AB}^{\mathrm{model}}(z)
  \ket{\Phi_{\mathrm{HF}}^{AB}}
  =E_{\mathrm{HF}}^{AB}(z).
  \label{eq:deep_qsci_hf_matching_condition}
\end{equation}
The correction in Eq.~\eqref{eq:deep_qsci_hf_reference_correction} incorporates the state-independent difference between the retained model terms and the full-dimer HF reference. This difference includes the internuclear repulsion, frozen-core contributions, and other omitted effects at the HF level. The correction is not fitted to a correlated energy or to a reference interaction energy. Because it depends on $z$, it contributes directly to the calculated interaction-energy curve and is not merely a global choice of the energy origin.

\subsection{local QSCI}
\label{subsec:deep_qsci_local_spaces}

We perform QSCI for the isolated-monomer Hamiltonian
$\hat H_{\mathrm{mono}}^{\mathrm{act}}$ as described in Sec.~\ref{subsec:qsci}.  The QSCI output state is
\begin{equation}
  \ket{\psi_0}
  =\sum_{\ket{x}\in\mathcal S_R}c_x\ket{x},
  \label{eq:deep_qsci_monomer_reference}
\end{equation}
where $\mathcal S_R$ is the set of selected Slater determinants represented by computational-basis states. The same coefficients and determinant occupations are used for the two identical monomers,
\begin{equation}
  \ket{\psi_0^{(i)}}
  =\sum_{\ket{x}\in\mathcal S_R}c_x\ket{x^{(i)}},
  \label{eq:deep_qsci_identical_references}
\end{equation}
where the two copies act on their respective monomer registers and $i\in\{A,B\}$.

Candidate local states are generated using spin-conserving one-body operators,
\begin{equation}
  \hat O_{pq\sigma}^{(i)}
  =\hat a_{p\sigma,i}^{\dagger}\hat a_{q\sigma,i},
  \label{eq:deep_qsci_local_excitation_operator}
\end{equation}
together with the identity operator.  If $m$ labels the selected operators, the nonorthogonal candidate states are
\begin{equation}
  \ket{\psi_m^{(i)}}
  =\hat O_m^{(i)}\ket{\psi_0^{(i)}}.
  \label{eq:deep_qsci_candidate_local_states}
\end{equation}
These operators preserve $N_\alpha$ and $N_\beta$ separately. The local basis therefore remains in the neutral charge sector of each monomer.

The overlap matrix of the candidate states is
\begin{equation}
  S_{mn}^{(i)}
  =\left\langle\psi_m^{(i)}\middle|\psi_n^{(i)}\right\rangle.
  \label{eq:deep_qsci_local_overlap_matrix}
\end{equation}
After linearly dependent or numerically unstable directions are removed, an orthonormal basis is constructed as
\begin{equation}
  \ket{\tilde\psi_a^{(i)}}
  =\sum_m P_{am}^{(i)}\ket{\psi_m^{(i)}},
  \qquad
  P^{(i)}S^{(i)}P^{(i)\dagger}=I.
  \label{eq:deep_qsci_orthonormal_local_basis}
\end{equation}
We denote the retained local space and its dimension by
\begin{equation}
  \mathcal K_i
  =\operatorname{span}
  \left\{\ket{\tilde\psi_a^{(i)}}\right\}_{a=1}^{K_i},
  \qquad
  \dim\mathcal K_i=K_i.
  \label{eq:deep_qsci_local_space}
\end{equation}

Because every state in $\mathcal K_A\otimes\mathcal K_B$ has fixed electron numbers on both monomers, the projected hopping matrix elements vanish,
\begin{equation}
  \bra{\tilde\psi_a^{(A)}}
  \bra{\tilde\psi_b^{(B)}}
  \hat a_{p\sigma,A}^{\dagger}\hat a_{q\sigma,B}
  \ket{\tilde\psi_c^{(A)}}
  \ket{\tilde\psi_d^{(B)}}
  =0.
  \label{eq:deep_qsci_vanishing_hopping_matrix_element}
\end{equation}
Thus, the hopping terms are retained in the general model interaction but do not contribute in the particle-number-conserving calculations reported here.  Charge transfer could be included by adding local states from the  $N_e+1$ and $N_e-1$ sectors while fixing only the total number of electrons in the dimer. Such sectors are not included in the present calculations.

\subsection{Effective Hamiltonian}
\label{subsec:deep_qsci_effective_problem}

The reduced dimer space is
\begin{equation}
  \mathcal K_{AB}=\mathcal K_A\otimes\mathcal K_B.
  \label{eq:deep_qsci_product_space}
\end{equation}
In the orthonormal product basis, the effective-Hamiltonian matrix elements are
\begin{align}
  (H_{\mathrm{eff}})_{ab,cd}
  &=\bra{\tilde\psi_a^{(A)}}\bra{\tilde\psi_b^{(B)}}
  \hat H_{AB}^{\mathrm{model}}
  \ket{\tilde\psi_c^{(A)}}\ket{\tilde\psi_d^{(B)}} \notag\\
  &=\bra{\tilde\psi_a^{(A)}}
  \hat H_A^{\mathrm{act}}
  \ket{\tilde\psi_c^{(A)}}\delta_{bd} \notag\\
  &\quad+\delta_{ac}
  \bra{\tilde\psi_b^{(B)}}
  \hat H_B^{\mathrm{act}}
  \ket{\tilde\psi_d^{(B)}} \notag\\
  &\quad+\bra{\tilde\psi_a^{(A)}}\bra{\tilde\psi_b^{(B)}}
  \hat V_{AB}^{\mathrm{model}}
  \ket{\tilde\psi_c^{(A)}}\ket{\tilde\psi_d^{(B)}}
  \label{eq:deep_qsci_effective_hamiltonian_matrix}
\end{align}
The dependence on $z$ is suppressed in Eq.~\eqref{eq:deep_qsci_effective_hamiltonian_matrix} for clarity. The matrix dimension is $K_AK_B$, or $K^2$ when the two identical monomers use the same local dimension $K$.

In the implementation, $\hat V_{AB}^{\mathrm{model}}$ is mapped to Pauli operators using the Jordan--Wigner transformation.  This mapping can produce terms acting on only one monomer and constant terms even when the original fermionic operator couples the two monomers.  For example, for spin orbitals $p$ on $A$ and $r$ on $B$, $\hat n_{p,A}\hat n_{r,B}=\frac{1}{4}\left(I-\hat Z_{p,A}-\hat Z_{r,B}+\hat Z_{p,A}\hat Z_{r,B}\right).$  All four terms are retained as parts of the same intermonomer operator.  In assembling $H_{\mathrm{eff}}$, the single-monomer-support terms are included once in the corresponding effective local block, the two-monomer terms are included in the coupling block, and the constant is added once to the diagonal.  They are therefore neither discarded nor independently introduced or double counted as additional physical interactions.

The effective Hamiltonian is diagonalized classically,
\begin{equation}
  H_{\mathrm{eff}}\boldsymbol C^{(n)}
  =E_n^{\mathrm{eff}}\boldsymbol C^{(n)}.
  \label{eq:deep_qsci_effective_eigenvalue_problem}
\end{equation}
The Deep QSCI dimer energy is its lowest eigenvalue,
\begin{equation}
  E_{AB}^{\mathrm{Deep\,QSCI}}(z)
  =\min_n E_n^{\mathrm{eff}}(z),
  \label{eq:deep_qsci_dimer_energy}
\end{equation}
and the corresponding state is
\begin{equation}
  \ket{\Psi_{AB}^{\mathrm{Deep\,QSCI}}}
  =\sum_{a=1}^{K_A}\sum_{b=1}^{K_B}
  C_{ab}^{(0)}
  \ket{\tilde\psi_a^{(A)}}
  \otimes\ket{\tilde\psi_b^{(B)}}.
  \label{eq:deep_qsci_dimer_state}
\end{equation}
For the model Hamiltonian defined in Eq.~\eqref{eq:deep_qsci_model_hamiltonian}, the Rayleigh--Ritz principle gives $E_0^{\mathrm{model}}(z)\leq E_{AB}^{\mathrm{Deep\,QSCI}}(z)$, where $E_0^{\mathrm{model}}(z)$ is the exact ground-state energy of $\hat H_{AB}^{\mathrm{model}}(z)$ in the symmetry sector considered. This bound applies only to the model Hamiltonian.  It does not provide a variational bound to the ground-state energy of the full dimer active-space Hamiltonian.

Finally, the interaction energy is evaluated as
\begin{equation}
  E_{\mathrm{int}}^{\mathrm{Deep\,QSCI}}(z)
  =E_{AB}^{\mathrm{Deep\,QSCI}}(z)
  -2E_{\mathrm{mono}}^{\mathrm{QSCI}},
  \label{eq:deep_qsci_interaction_energy}
\end{equation}
where $E_{\mathrm{mono}}^{\mathrm{QSCI}}$ is obtained for an isolated monomer without ghost basis functions.  

To separate the contribution fixed by the HF reference correction from the energy lowering obtained with Deep QSCI, we define the monomer active-space HF reference energy as
\begin{equation}
  E_{\mathrm{mono}}^{\mathrm{HF,act}}
  =\bra{\Phi_{\mathrm{HF}}^A}
  \hat H_{\mathrm{mono}}^{\mathrm{act}}
  \ket{\Phi_{\mathrm{HF}}^A}.
  \label{eq:deep_qsci_monomer_hf_reference}
\end{equation}
Equation~\eqref{eq:deep_qsci_interaction_energy} can then be written exactly as
\begin{equation}
  E_{\mathrm{int}}^{\mathrm{Deep\,QSCI}}(z)
  =E_{\mathrm{int}}^{\mathrm{HF,ref}}(z)
  +\Delta E_{\mathrm{int}}^{\mathrm{corr}}(z),
  \label{eq:deep_qsci_interaction_energy_decomposition}
\end{equation}
where $E_{\mathrm{int}}^{\mathrm{HF,ref}}(z)=E_{\mathrm{HF}}^{AB}(z)-2E_{\mathrm{mono}}^{\mathrm{HF,act}}$ is the interaction energy associated with the matched HF reference, and
\begin{align}
  \Delta E_{\mathrm{int}}^{\mathrm{corr}}(z)
  &=\left(E_{AB}^{\mathrm{Deep\,QSCI}}(z)
  -E_{\mathrm{HF}}^{AB}(z)\right)
  \notag \\
  &\quad-2\left(E_{\mathrm{mono}}^{\mathrm{QSCI}}
  -E_{\mathrm{mono}}^{\mathrm{HF,act}}\right)
  \label{eq:deep_qsci_correlation_interaction_energy}
\end{align}
is the additional interaction contribution obtained beyond this HF reference.  Equation~\eqref{eq:deep_qsci_interaction_energy_decomposition} is an algebraic identity and does not introduce a further approximation. The quantity
$\Delta E_{\mathrm{int}}^{\mathrm{corr}}(z)$ isolates the contribution obtained from the Deep QSCI treatment after the geometry-dependent HF matching has been removed. It should not be interpreted as the exact correlation contribution of the full dimer Hamiltonian because it is evaluated with the model
Hamiltonian and the reduced local spaces. When $E_{\mathrm{mono}}^{\mathrm{HF,act}}$ equals the conventional isolated-monomer HF energy, $E_{\mathrm{int}}^{\mathrm{HF,ref}}(z)$ is the conventional uncorrected HF interaction energy.

Because $E_{\mathrm{mono}}^{\mathrm{QSCI}}$ is evaluated without ghost basis functions, Eq.~\eqref{eq:deep_qsci_interaction_energy} defines an uncorrected interaction energy that may contain basis-set superposition error. Comparisons with counterpoise-corrected reference data must account for this difference.

The present calculation involves approximations associated with the QSCI-selected determinant space, the local excitation space, the restricted form of $\hat V_{AB}^{\mathrm{model}}$, the index mapping between the isolated and L\"owdin-orthogonalized orbitals, and the exclusion of charge-transfer sectors.  The geometry-dependent HF reference correction is also part of the model definition.  The quantum calculation is confined to one monomer QSCI workflow for each basis-set and active-space choice, while the dimer effective Hamiltonian is assembled and diagonalized classically.

\section{Experimental setup}
\label{sec:computational_setup}
\begin{table*}[t]
\centering
\caption{
Computational resources used in the present Deep QSCI calculations.  The values were extracted from the actual Löwdin-orthogonalized datasets and the corresponding Deep QSCI results used in this study.  The active-space notation refers to the entire dimer; $\mathrm{CAS}(12e,12o)$ and $\mathrm{CAS}(28e,20o)$ are constructed from two monomer spaces of $\mathrm{CAS}(6e,6o)$ and $\mathrm{CAS}(14e,10o)$, respectively.
The number of Pauli terms denotes that of the monomer Hamiltonian used in the QSCI calculation.  The monomer input states for these Deep QSCI calculations are prepared using UCCSD.
}
\setlength{\tabcolsep}{5pt}
\resizebox{\textwidth}{!}{
\begin{tabular}{lccccccccc}
\hline\hline
Active space
& Basis set
& Monomer qubits
& Dimer qubits
& Monomer Pauli terms
& Config. size
& Shots per $\theta$
& $K_A$
& $K_B$
& $\dim(H_{\mathrm{eff}})$ \\
\hline
$\mathrm{CAS}(12e,12o)$ & 6-31G** & 12 & 24 & 239  & 50 & 10000 & 44 & 44 & 1936 \\
$\mathrm{CAS}(12e,12o)$ & cc-pVDZ & 12 & 24 & 239  & 50 & 10000 & 44 & 44 & 1936 \\
\hline
$\mathrm{CAS}(28e,20o)$ & 6-31G** & 20 & 40 & 1751 & 50 & 10000 & 93 & 93 & 8649 \\
$\mathrm{CAS}(28e,20o)$ & cc-pVDZ & 20 & 40 & 1751 & 50 & 10000 & 70 & 70 & 4900 \\
\hline\hline
\end{tabular}
}
\label{tab:deepqsci_resources}
\end{table*}

We apply Deep QSCI to the benzene dimer in the sandwich configuration.  Each monomer is described using a rigid benzene geometry with C--C and C--H bond lengths of $r_{\mathrm{CC}}=1.390~\text{\AA}$ and $r_{\mathrm{CH}}=1.080~\text{\AA}$, respectively. The monomer geometry is not reoptimized as a function of the intermolecular separation.  Monomer $B$ is obtained by rigidly translating monomer $A$ along the stacking axis.  The Deep QSCI results reported here cover intermolecular separation from $z=3.0$ to $6.0~\text{\AA}$.

We consider two active spaces for each monomer.  The first is a $\pi$-electron active space, $\mathrm{CAS}(6e,6o)$, consisting of the six $\pi$ orbitals derived from the carbon $2p_z$ manifold.  The second is an extended active space, $\mathrm{CAS}(14e,10o)$, obtained by augmenting the $\pi$ space with eight additional valence electrons in four orbitals near the Fermi level.  Combining the two identical monomer active spaces without further truncation yields the corresponding dimer spaces $\mathrm{CAS}(12e,12o)$ and $\mathrm{CAS}(28e,20o)$, respectively, as listed in Table~\ref{tab:deepqsci_resources}.  Calculations are performed using the 6-31G** and cc-pVDZ basis sets.

For each monomer, the active-space one- and two-electron integrals are generated from a restricted Hartree--Fock reference using PySCF~\cite{Sun2020PySCF} (v2.13.0) through the OpenFermion--PySCF interface (v0.5).  The resulting fermionic Hamiltonian is mapped onto qubits using the Jordan--Wigner transformation implemented in OpenFermion~\cite{McClean2020OpenFermion} (v1.7.1).  The number of Pauli terms in each monomer qubit Hamiltonian is reported in Table~\ref{tab:deepqsci_resources}.  As described in Sec.~\ref{sec:deep_qsci}, only the monomer Hamiltonians are used in the quantum part of the calculation, whereas the intermolecular contribution is constructed and treated classically.

For each basis-set and active-space choice, the monomer QSCI configuration space used in Deep QSCI is generated using a unitary coupled-cluster ansatz containing single and double excitations (UCCSD), initialized in the HF reference state. The corresponding time-evolution operator is approximated using a first-order Suzuki--Trotter decomposition, with all UCC amplitudes assigned a common value $\theta$. Instead of variationally optimizing the amplitudes independently, we perform a grid search over
$\theta\in\{0,0.05,0.1,0.15,0.2\}$ and select the value that yields the lowest QSCI subspace ground-state energy.

The direct dimer QSCI result included for comparison is obtained using UCCS input-state preparation.  UCCS was adopted because constructing and simulating a UCCSD circuit for the full dimer produces a prohibitively deep circuit and requires substantially greater computational time.  This choice is therefore an implementation constraint imposed by the computational cost of the direct dimer calculation, rather than an indication that UCCS is intrinsically more accurate.  Consequently, the direct QSCI and Deep QSCI results differ not only in whether the dimer is partitioned into monomers, but also in the excitation rank used in the state-preparation ansatz.  Their comparison should therefore be interpreted as a comparison of the computationally feasible workflows implemented here rather than as an isolated test of the hierarchical reduction alone.

Quantum circuits are constructed using the Classiq SDK~\cite{minerbi2022quantum, vax2025qmod, goldfriend2024design} (v1.25.0) and evaluated using noiseless simulation followed by sampling with $10^4$ shots for each value of $\theta$. Thus, one monomer QSCI workflow uses $5\times10^4$ shots in total for the five-point parameter scan. For each $\theta$, the $R=50$ most frequently sampled electron configurations are retained. The Hamiltonian is then projected onto the resulting QSCI subspace and diagonalized classically. This procedure yields the monomer reference state $\ket{\psi_0}$ defined in Eq.~\eqref{eq:deep_qsci_monomer_reference}.

The complete monomer QSCI workflow, including the scan over $\theta$, is performed once for each basis set and active space. Its selected determinants and coefficients are reused for both monomers throughout the separation scan. Consequently, only the geometry-dependent construction of $\hat V_{AB}^{\mathrm{model}}$, assembly of $H_{\mathrm{eff}}$, and classical diagonalization are repeated at each $z$.

Independent CASCI and CCSD(T) calculations are included as external reference curves.  The CASCI calculation uses $\mathrm{CAS}(28e,20o)$ with 6-31G**, whereas no CAS truncation is assigned to the CCSD(T) calculations.  Their tabulated interaction energies use counterpoise-corrected monomer references.  They do not evaluate the Deep QSCI model Hamiltonian in Eq.~\eqref{eq:deep_qsci_model_hamiltonian}, and their frozen-orbital convention is not assumed to coincide with the active/frozen partition of the model.  The comparison is therefore used to assess qualitative curve shape and energy scale, rather than as a matched-Hamiltonian variational benchmark.

For each monomer $i$, the local space $\mathcal K_i$ introduced in Sec.~\ref{subsec:deep_qsci_local_spaces} is constructed by applying the identity and all one-body operators $\hat a_{p\sigma}^{\dagger}\hat a_{q\sigma}$ involving the selected boundary spin orbitals to $\ket{\psi_0^{(i)}}$.  In the calculations reported here, all active spin orbitals are selected as boundary orbitals.  The operators preserve $N_\alpha$ and $N_\beta$ separately; no claim of exact $S^2$ adaptation is made.  Linear dependencies and numerically unstable directions are removed by diagonalizing the corresponding overlap matrix and retaining only eigenvectors with eigenvalues larger than $10^{-6}$. The local basis is restricted to the particle-number sector corresponding to the neutral monomer.

The resulting local-space dimensions $K_A$ and $K_B$, together with the effective-Hamiltonian dimension
$\dim(H_{\mathrm{eff}})=K_AK_B$, given by Eq.~\eqref{eq:deep_qsci_effective_hamiltonian_matrix}, are summarized in Table~\ref{tab:deepqsci_resources} for each active space and basis set. Finally, $H_{\mathrm{eff}}$ is assembled and diagonalized classically to obtain the Deep QSCI dimer energy defined in Eq.~\eqref{eq:deep_qsci_dimer_energy}.

\section{Results}
\label{sec:result}
\begin{figure}
    \centering
    \includegraphics[width=1.0\linewidth]{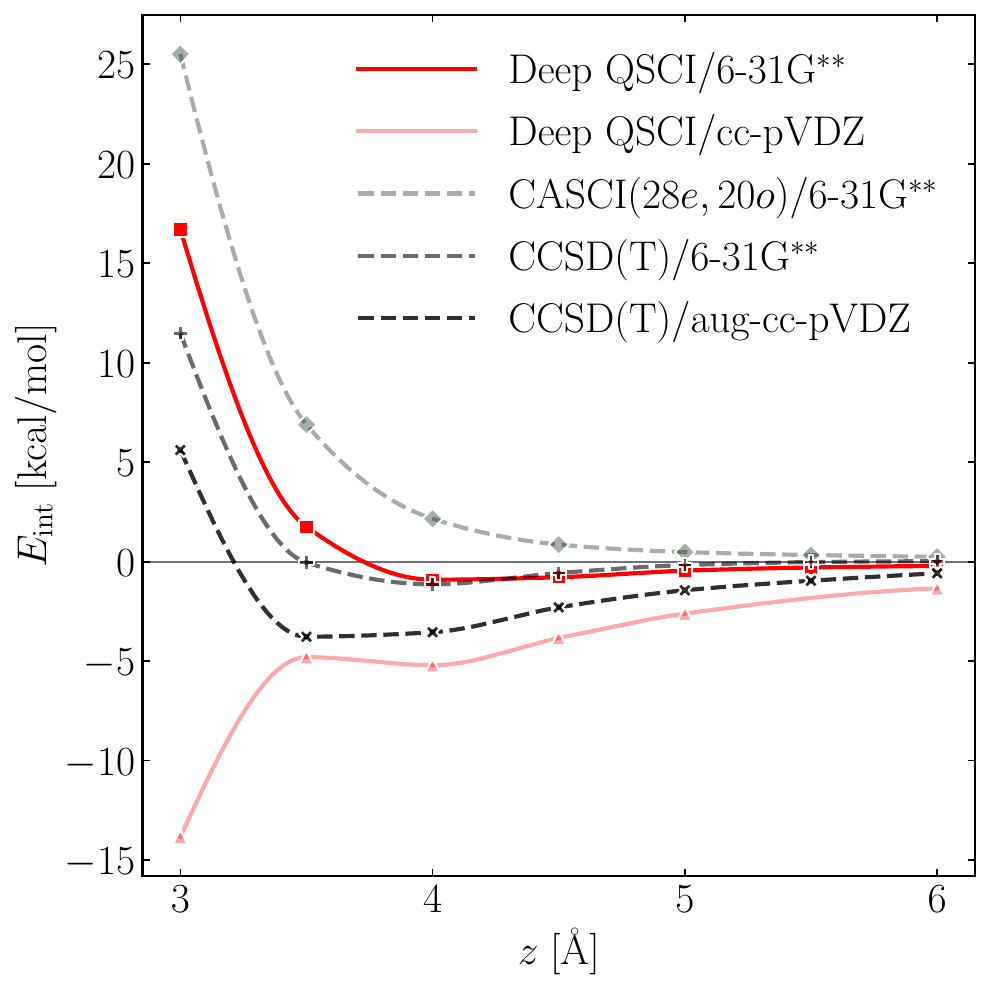}
    \caption{Potential-energy curves of the sandwich benzene dimer as functions of the intermolecular separation $z$.  (a) Absolute dimer energies obtained with Deep QSCI, CASCI, and CCSD(T); the inset shows $E_{AB}(z)-E_{AB}(6.0~\text{\AA})$.  (b) Interaction energies.  The plotted Deep QSCI and CASCI calculations use the dimer space $\mathrm{CAS}(28e,20o)$; no CAS truncation is assigned to CCSD(T).  The Deep QSCI values are defined relative to isolated monomers without ghost functions, whereas the classical CASCI and CCSD(T) interaction energies are counterpoise corrected.  Symbols denote the calculated data points, and the shape-preserving spline curves are guides to the eye.  Basis sets are indicated in the legend.}
    \label{fig:dimer_and_interaction_energy_ext_pi}
\end{figure}

\begin{table*}[tb]
    \centering
    \caption{Interaction energies (kcal/mol) used in the present comparison.  The direct QSCI calculation uses UCCS, whereas the monomer QSCI step in Deep QSCI uses UCCSD.  Deep QSCI and direct QSCI use isolated-monomer references without ghost functions and are therefore not counterpoise corrected.  The classical CASCI and CCSD(T) values are counterpoise corrected.  The direct-QSCI row retains a large positive offset at $6.0$~\AA and is consequently reported only as a diagnostic result.}
    \begin{tabular}{lllrrrrrr}\hline\hline
        Active space 
        & Method 
        & Basis set 
        & $z=3.0$\AA 
        & $z=3.5$\AA
        & $z=4.0$\AA
        & $z=4.5$\AA
        & $z=5.0$\AA
        & $z=6.0$\AA\\\hline
        $\mathrm{CAS}(12e,12o)$
        & QSCI
        & 6-31G**
        & $50.874600$
        & $30.675000$
        & $25.845300$
        & $25.035800$
        & $24.793500$
        & $24.557500$\\
        $\mathrm{CAS}(12e,12o)$
        & Deep QSCI
        & 6-31G**
        & $21.825877$
        & $3.7956315$
        & $0.10159449$
        & $-0.18831960$
        & $-0.055910677$
        & $-0.0054852712$\\
        $\mathrm{CAS}(12e,12o)$
        & Deep QSCI
        & cc-pVDZ
        & $22.446408$
        & $4.3527634$
        & $0.47749518$
        & $-0.14493069$
        & $-0.093117695$
        & $-0.014293905$\\\hline
        $\mathrm{CAS}(28e,20o)$
        & Deep QSCI 
        & 6-31G**
        & $16.698166$
        & $1.7495162$
        & $-0.90769539$
        & $-0.78227530$
        & $-0.44978479$
        & $-0.21575607$\\
        $\mathrm{CAS}(28e,20o)$
        & CASCI
        & 6-31G**
        & $25.509887$
        & $6.8912420$
        & $2.1610820$
        & $0.85987200$
        & $0.48210700$
        & $0.24682700$\\
        --
        & CCSD(T)
        & 6-31G**
        &$11.485287$
        &$-0.046593000$
        &$-1.1391300$
        &$-0.56682900$
        &$-0.16748900$
        &$0.025684000$\\
        $\mathrm{CAS}(28e,20o)$
        & Deep QSCI 
        & cc-pVDZ
        & $-13.828069$
        & $-4.7954092$
        & $-5.2050916$
        & $-3.8350583$
        & $-2.6090980$
        & $-1.3418349$\\
        --
        & CCSD(T)
        & aug-cc-pVDZ
        &$5.6128900$
        &$-3.7695870$
        &$-3.5462570$
        &$-2.2953730$
        &$-1.4363820$
        &$-0.58267400$\\\hline\hline
    \end{tabular}
    \label{tab:interaction-energy}
\end{table*}

\begin{table}[tb]
    \centering
    \caption{Decomposition of the extended-space Deep QSCI interaction energy obtained with 6-31G** (kcal/mol).  $E_{\mathrm{int}}^{\mathrm{HF,ref}}$ and $\Delta E_{\mathrm{int}}^{\mathrm{corr}}$ are defined in Eqs.~\eqref{eq:deep_qsci_interaction_energy_decomposition} and \eqref{eq:deep_qsci_correlation_interaction_energy}.  $E_{\mathrm{int}}^{\mathrm{prod}}$ is the QSCI product-state expectation value relative to $2E_{\mathrm{mono}}^{\mathrm{QSCI}}$, and $\Delta E_{\mathrm{diag}}=E_{\mathrm{int}}^{\mathrm{Deep\,QSCI}}-E_{\mathrm{int}}^{\mathrm{prod}}$.}
    \begin{tabular}{crrrrr}\hline\hline
        $z$ (\AA) & $E_{\mathrm{int}}^{\mathrm{HF,ref}}$
        & $\Delta E_{\mathrm{int}}^{\mathrm{corr}}$
        & $E_{\mathrm{int}}^{\mathrm{prod}}$
        & $\Delta E_{\mathrm{diag}}$
        & $E_{\mathrm{int}}^{\mathrm{Deep\,QSCI}}$\\\hline
        $3.0$ & $26.620$ & $-9.922$ & $26.532$ & $-9.834$ & $16.698$\\
        $3.5$ & $ 6.420$ & $-4.671$ & $ 6.354$ & $-4.605$ & $ 1.750$\\
        $4.0$ & $ 1.635$ & $-2.542$ & $ 1.584$ & $-2.491$ & $-0.908$\\
        $4.5$ & $ 0.780$ & $-1.562$ & $ 0.740$ & $-1.522$ & $-0.782$\\
        $5.0$ & $ 0.597$ & $-1.047$ & $ 0.566$ & $-1.015$ & $-0.450$\\
        $6.0$ & $ 0.338$ & $-0.554$ & $ 0.317$ & $-0.533$ & $-0.216$\\\hline\hline
    \end{tabular}
    \label{tab:interaction-decomposition}
\end{table}

We calculate the isolated benzene-monomer QSCI ground-state energy $E_{\mathrm{mono}}^{\mathrm{QSCI}}$ for each active-space.  For the monomer $\mathrm{CAS}(6e,6o)$ space, the obtained energies are $-230.736631$ Ha with 6-31G** and $-230.744611$ Ha with cc-pVDZ.  For the extended monomer $\mathrm{CAS}(14e,10o)$ space, the corresponding energies were $-230.713734$ Ha and $-230.730871$ Ha, respectively.  Each monomer energy was reused as the reference at all intermolecular separation, and the Deep QSCI interaction energy was evaluated as Eq.~\eqref{eq:deep_qsci_interaction_energy}.  The resulting interaction energies are listed in Table~\ref{tab:interaction-energy}. The absolute ground-state energies of the benzene dimer are provided in Appendix~\ref{sec:dimer data}.

In the $\rm CAS(12e,12o)$ space, Deep QSCI yields shallow minima near $z=4.5~\text{\AA}$.  The minimum interaction energies are $-0.188$ kcal/mol with 6-31G** and $-0.145$ kcal/mol with cc-pVDZ.  By comparison, direct QSCI with $\mathrm{CAS}(12e,12o)$ and 6-31G** gives positive interaction energies throughout the reported range.  The interaction energy remains $+24.56$ kcal/mol at $z=6.0~\text{\AA}$.  This large residual value limits a quantitative interpretation of the direct QSCI results and may reflect limitations in the prepared state or the selected determinant space. The present data do not identify the cause.  The two calculations also use different state-preparation ansatzes.  Direct QSCI uses UCCS, whereas the monomer QSCI calculation in Deep QSCI uses UCCSD. Comparison with extended-space Deep QSCI further introduces a change in the active space. These comparisons therefore do not separate the effects of the ansatz, the active space, and the Deep QSCI construction.

For $\rm CAS(28e,20o)$ space, the interaction energy of Deep QSCI with 6-31G** changes sign between $z=3.5$ and $4.0~\text{\AA}$ and reaches $-0.908$ kcal/mol at $z=4.0~\text{\AA}$.  Table~\ref{tab:interaction-decomposition} separates the contributions to this attraction. At this separation, the matched HF reference gives a repulsive interaction of $E_{\mathrm{int}}^{\mathrm{HF,ref}}=+1.635$ kcal/mol. The product of the two monomer QSCI states also gives a repulsive interaction of $E_{\mathrm{int}}^{\mathrm{prod}}=+1.584$ kcal/mol.  Diagonalization in the reduced product space lowers the energy by $\Delta E_{\mathrm{diag}}=-2.491$ kcal/mol and gives the final interaction energy of $-0.908$ kcal/mol.  Within the present model, the attraction therefore arises from mixing the local basis states through diagonalization. Neither the HF reference correction nor the monomer QSCI product state alone yields an attractive interaction at this separation.

The extended-space Deep QSCI result with 6-31G** is close to the corresponding CCSD(T) result near the minimum.  At $z=4.0~\text{\AA}$, the interaction energies are $-0.908$ and $-1.139$ kcal/mol, respectively. However, the two calculations use different treatments of basis-set superposition error. The Deep QSCI result is uncorrected, whereas the tabulated CCSD(T) result includes counterpoise correction. This numerical agreement therefore does not establish quantitative accuracy. Differences also remain in the position and depth of the minimum.

The CASCI interaction energy with 6-31G** remains positive throughout the reported range. It decreases from $25.51$ kcal/mol at $z=3.0~\text{\AA}$ to $0.247$ kcal/mol at $z=6.0~\text{\AA}$. CASCI and Deep QSCI use different Hamiltonian constructions, so this comparison does not establish an accuracy advantage of Deep QSCI over CASCI. It shows that the present Deep QSCI model yields an attractive region under the chosen conditions.

For extended-space Deep QSCI with cc-pVDZ, the energy decomposition shows a large cancellation. At $z=4.0~\text{\AA}$, the monomer QSCI product state gives an interaction energy of $+166.36$ kcal/mol. Subsequent diagonalization lowers the energy by $171.56$ kcal/mol, giving a final interaction energy of $-5.21$ kcal/mol. Cancellations exceeding $170$ kcal/mol persist even at $z=6.0~\text{\AA}$. The final interaction energy is thus a small difference between two large contributions. This behavior may involve limitations of the monomer QSCI state, the orbital-index mapping, or the reduced-space construction. The present results do not distinguish these possibilities. We therefore do not interpret the extended-space cc-pVDZ curve as a quantitative prediction of the benzene-dimer interaction.

Overall, the 6-31G** results demonstrate that a model interaction curve with an attractive region can be obtained by reusing a single 20-qubit monomer QSCI workflow for both monomers and across separation.  Quantitative accuracy for the physical benzene-dimer interaction remains to be established. Further assessment of the monomer QSCI states, orbital consistency, and reduced-space convergence is needed to clarify the basis-set dependence while preserving the computational benefit of monomer reuse.

\section{Discussion}
\label{sec:discussion}

The main resource benefit of Deep QSCI is the reduction in quantum-register size. In the encoding used here, direct QSCI for the dimer space $\mathrm{CAS}(28e,20o)$ requires $40$ qubits, whereas Deep QSCI uses $20$ qubits for the monomer space $\mathrm{CAS}(14e,10o)$. For the identical rigid monomers considered here, the same monomer QSCI result is reused for both monomers at every intermolecular separation. The monomer quantum sampling therefore does not need to be repeated during the separation scan. These benefits come at the classical cost of constructing and diagonalizing an effective Hamiltonian of dimension $K_AK_B$. Since computation times and scaling were not benchmarked, the present results do not establish an overall computational speedup or quantum advantage. They instead demonstrate a reduction in quantum-register size and in the need for repeated quantum sampling for an active space that remains challenging to treat directly on current NISQ devices.

For 6-31G**, both the matched HF reference and the product of the two monomer QSCI states give repulsive interaction energies at $z=4.0$~\AA. The interaction becomes attractive only after diagonalization in the reduced product space, as shown in Table~\ref{tab:interaction-decomposition}. Within the present model, mixing the local basis states therefore provides the additional stabilization required to produce an attractive interaction. This result is qualitatively consistent with the established importance of electron correlation in benzene-dimer binding~\cite{tsuzuki2002origin}. However, the resulting energy lowering cannot be identified solely with the physical dispersion energy because its magnitude is also affected by approximations in the model Hamiltonian, including the restricted intermonomer interaction and the construction of the reduced local basis.

Although the absolute Deep QSCI dimer energies are higher than the corresponding CASCI energies, the 6-31G** Deep QSCI interaction energies are closer to the available CCSD(T) values in the attractive region. This agreement does not establish that Deep QSCI is more accurate than CASCI. An interaction energy is a difference between dimer and monomer energies, so its accuracy depends on how errors in these energies cancel. Moreover, Deep QSCI and CASCI use different Hamiltonian constructions. The Deep QSCI interaction energies are also uncorrected for basis-set superposition error, whereas the tabulated CCSD(T) values include counterpoise correction. The agreement with CCSD(T) is therefore encouraging, but it does not provide a controlled comparison of accuracy.

Reusing the isolated-monomer Hamiltonian allows a single monomer QSCI calculation to provide the local basis for both monomers throughout the separation scan. To preserve this benefit, the monomer Hamiltonians are not reconstructed after joint L"owdin orthogonalization. Instead, the orthogonalized embedded orbitals used to evaluate $\hat V_{AB}^{\mathrm{model}}$ are associated with the original monomer orbitals through their orbital indices. The resulting Hamiltonian therefore combines monomer and interaction terms expressed in different orbital representations. This index mapping is a deliberate modeling approximation introduced to retain monomer reuse. The large cancellation between the product-state interaction energy and the energy lowering obtained with cc-pVDZ highlights the need to assess this approximation, although the cancellation alone does not identify its cause.

A further route to improving the method is to expand the local basis used to construct the effective Hamiltonian. Previous Deep VQE studies have shown that the choice of local subspace affects both accuracy and resource requirements~\cite{Fujii2022DeepVQE,Mizuta2021DeepVQEexcited,Erhart2022LocalBases}. Within the present fixed-particle-number construction, convergence can first be assessed by increasing the selected determinant space and expanding the set of local excitation operators. This would help distinguish limitations of the reduced space from those of the model Hamiltonian.

The local basis could also be extended beyond the fixed electron number used for each monomer. In addition to the neutral sector $(N_A,N_B)$, the effective space could include $(N_A+1,N_B-1)$ and $(N_A-1,N_B+1)$ while preserving the total electron number. These sectors would allow explicit intermonomer charge-transfer configurations, which are excluded from the present local basis. For identical monomers, the local basis for each electron-number sector could still be reused for both monomers. However, this extension would increase the effective-space dimension and require additional local states and matrix elements between sectors. A common orbital set would simplify these calculations, but might describe charged monomer states less accurately than orbitals optimized separately for each sector. Separate orbital optimization would, in turn, complicate the evaluation of matrix elements between the local states. Developing this extension while maintaining orbital consistency and a fixed total electron number remains a direction for future work.


\section{Conclusion}
\label{sec:conclusion}

We proposed Deep QSCI, a divide-and-conquer extension of quantum-selected configuration interaction for evaluating a model interaction-energy curve of the benzene dimer. By decomposing the dimer into two monomers, the quantum calculation for the extended active space is reduced from the dimer space $\mathrm{CAS}(28e,20o)$, which requires $40$ qubits, to the monomer space $\mathrm{CAS}(14e,10o)$, which requires $20$ qubits.  A single isolated-monomer QSCI workflow is reused for both monomers and all separation for each basis-set and active-space choice.  The dimer energy is then evaluated by constructing an effective Hamiltonian from reduced local bases generated around the monomer QSCI state and the model intermonomer interaction, followed by classical diagonalization.

With 6-31G**, diagonalization in the reduced product space yields an attractive region around $z\simeq4.0$--$4.5$~\AA that is absent from both the matched HF reference and the QSCI product state.  Its energy scale near $z=4.0$~\AA is close to the available CCSD(T)/6-31G** result.  This agreement is encouraging, but the distinct Hamiltonian constructions and counterpoise treatments prevent it from serving as a quantitative validation.

The extended-space cc-pVDZ calculation exhibits a cancellation of more than $170$ kcal/mol between the QSCI product-state interaction and the diagonalization lowering, and is therefore treated as a diagnostic failure rather than a quantitative interaction energy.  These results demonstrate the reuse of a single monomer QSCI calculation to obtain a model dimer curve with a smaller quantum register; quantitative accuracy for the physical benzene dimer remains to be established.  A consistent treatment of the orbital representations, tests of selected-space convergence, and comparison under matched counterpoise conditions are required next.

A further direction is to enlarge the local basis beyond the fixed particle-number sectors used in the present work so that charge-transfer configurations can be explicitly incorporated into the effective Hamiltonian.  Such an extension will require careful control of the effective-space dimension, conservation of the total particle number, and consistency of the orbital representation across different local particle-number sectors.

\section{Acknowledgement}
The authors thank Kazuki Matsushita, Masayuki Kobayashi, Hayato Kunugi, Tatsuo Akaki and Takahiro Hata for helpful discussions and for their comments on the manuscript.  The part of this work was supported by the "NEDO Challenge, Quantum Computing 'Solve Social Issues!'" program, conducted by the New Energy and Industrial Technology Development Organization (NEDO) and the Ministry of Economy, Trade and Industry, Japan.

\bibliographystyle{apsrev4-2}
\bibliography{main}
\clearpage
\onecolumngrid
\appendix

\section{Dimer data}
\label{sec:dimer data}
{\setlength{\tabcolsep}{5.5pt}
\begin{table*}[tb]
    \centering
    \caption{Dimer energies (Ha) used in the present comparison.  The direct QSCI calculation uses UCCS, whereas the monomer QSCI step in Deep QSCI uses UCCSD.  A dash in the active-space column indicates that no CAS truncation was assigned to the corresponding CCSD(T) calculation.  Because the methods employ different Hamiltonian constructions, ansatzes, and orbital treatments, the absolute energies should not be interpreted as a common variational sequence.}
    \begin{tabular}{lllrrrrr}\hline\hline
        Active space 
        & Method 
        & Basis set 
        & $z=3.0$\AA 
        & $z=3.5$\AA
        & $z=4.0$\AA
        & $z=4.5$\AA
        & $z=6.0$\AA\\ \hline
        $\mathrm{CAS}(12e,12o)$
        & QSCI 
        & 6-31G**
        & $-461.384449$   
        & $-461.416640$
        & $-461.424336$
        & $-461.425626$
        & $-461.426388$\\
        $\mathrm{CAS}(12e,12o)$
        & Deep QSCI 
        & 6-31G**
        & $-461.438479$
        & $-461.467212$
        & $-461.473099$
        & $-461.473561$
        & $-461.473269$\\
        $\mathrm{CAS}(12e,12o)$
        & Deep QSCI 
        & cc-pVDZ
        & $-461.453450$
        & $-461.482284$
        & $-461.488460$
        & $-461.489452$
        & $-461.489244$\\
        \hline
        $\mathrm{CAS}(28e,20o)$
        & Deep QSCI 
        & 6-31G**
        & $-461.400857$
        & $-461.424679$
        & $-461.428914$
        & $-461.428714$
        & $-461.427811$\\
        $\mathrm{CAS}(28e,20o)$
        & CASCI
        & 6-31G**
        & $-461.530704$
        & $-461.559415$
        & $-461.565836$
        & $-461.566879$
        & $-461.567313$\\
        --
        & CCSD(T)
        & 6-31G**
        & $-463.141431$
        & $-463.159808$
        & $-463.161550$
        & $-463.160637$
        & $-463.159693$\\
        $\mathrm{CAS}(28e,20o)$
        & Deep QSCI 
        & cc-pVDZ
        & $-461.483779$
        & $-461.469384$
        & $-461.470037$
        & $-461.467854$
        & $-461.463881$\\
        --
        & CCSD(T)
        & aug-cc-pVDZ
        & $-463.217595$
        & $-463.232546$
        & $-463.232191$
        & $-463.230197$
        & $-463.227468$\\\hline\hline
    \end{tabular}
    \label{tab:ground-energy-benzen-dimer}
\end{table*}
}

Table~\ref{tab:ground-energy-benzen-dimer} lists the benzene-dimer energies at selected intermolecular separations, including the direct QSCI and $\pi$-only Deep QSCI results omitted from the main figure. The $\pi$-only Deep QSCI energies reach their lowest tabulated values at $z=4.5$~\AA, whereas the extended-space Deep QSCI energies with 6-31G** reach a minimum at $z=4.0$~\AA. The extended-space cc-pVDZ results show a different distance dependence, as discussed in the main text. These data supplement the interaction-energy analysis and should not be used alone to rank the accuracy of the methods.

\end{document}